\documentclass[pss]{wiley2sp}

\usepackage{graphicx}
\usepackage{url}
\definecolor{copyrightcolor}{gray}{0.85}
\def\titlepageHeadlinewidth{\textwidth}
\journalname{\colorbox{copyrightcolor}{\small\parbox{\textwidth-2\fboxsep}{This is the peer reviewed version of the following article: R. Meyer, \textit{Vibrational band structure of nanoscale phononic crystals}, Phys. Status Solidi A \textbf{213}, No. 11, 2927–2935 (2016), which has been published in final form at \url{https://doi.org/10.1002/pssa.201600387}. This article may be used for non-commercial purposes in accordance with Wiley Terms and Conditions for Self-Archiving.}}}
\def\rightfootmark{}
\def\copyrightmark{}

\makeatletter
\def\@oddhead{\vbox to\headheight{\vss\baselineskip=3pt \hbox to\textwidth{\small\sffamily\the\authorrunning:~\the\titlerunning\hfill\thepage}\leavevmode\lower2pt\hbox to\textwidth{\hrulefill}}}
\makeatother

\title{Vibrational band structure of nanoscale phononic crystals} 
\titlerunning{Vibrational band structure of nanoscale phononic crystals}
\author{R. Meyer}
\authorrunning{R. Meyer}
\mail{rmeyer@cs.laurentian.ca}
\institute{%
{Department of Mathematics and Computer Science and Department of Physics,\\
 Laurentian University, 935 Ramsey Lake Road, Sudbury (Ontario), P3E 2C6, Canada}%
}

\abstract{%
The vibrational properties of two-dimensional phononic crystals are studied with large-scale molecular dynamics simulations and finite element method calculation. The vibrational band structure derived from the molecular dynamics simulations shows the existence of partial acoustic band gaps along the $\Gamma$-M direction. The band structure is in excellent agreement with the results from the finite element model, proving that molecular dynamics simulations can be used to study the vibrational properties of such complex systems. An analysis of the structure of the vibrational modes reveals how the acoustic modes deviate from the homogeneous bulk behaviour for shorter wavelengths and hints towards a decoupling of vibrations in the phononic crystal.
}

\begin{document}
\maketitle
%\pagestyle{myheadings}

%
%  Introduction
%
\section{Introduction}
Many promising technological applications in areas like ultrasonic imaging, wireless communications and energy harvesting are the driving force of the rising scientific interest in phononic crystals. Phononic crystals are periodically structured synthetic materials that use Bragg reflection to manipulate the propagation of pho\-nons and tailor the dispersion of elastic waves \cite{Maldovan:13a,Deymier:13a,Laude:15a,Khelif:16a}. Conceptually phononic crystals can be seen as an elastic analogue to photonic crystals \cite{Yablonovitch:87a,John:87a}  which use Bragg reflection in periodically structured systems to control the propagation of electromagnetic waves. 

The development of nanofabrication methods has enabled the synthesis of hypersonic phononic crystals with sub-micron periodicity lengths and operating frequencies in the GHz - THz range \cite{Gorishnyy:05a,Parsons:09a,Aliev:10a,Thomas:10a,Yu:10a,Hopkins:11a,Goettler:11a}. Recent theoretical and experimental studies have shown reduced thermal conductivities in such nanoscale phononic crystals \cite{Yu:10a,Hopkins:11a,Gillet:09a} making them candidates for thermoelectric materials with exceptionally high figures of merit $ZT$. This has generated interest in the properties of thermal phonons and their transport properties in nanoscale phononic crystals \cite{Davis:11a,Reinke:11a,Lacatena:15a,Nomura:15a,Nakagawa:15a}.

The calculation of vibrational properties --- in particular phonon band structures ---  started with the very first papers on phononic crystals  \cite{Sigalas:92a,Kushwaha:93a}. Several computational methods are available to study the phonon properties of phononic crystals \cite{Vasseur:13a}. Most of these methods are based on linear elasticity theory.  This becomes problematic for systems with structure sizes in the nanometer range. Macroscopic scaling laws and continuum theories reach their limits at the nanoscale and effects of the discrete atomic lattice become notable. Ramprasad and Shi \cite{Ramprasad:05a} have shown deviations of the phonon band structure from linear elasticity theory in nanoscale phononic crystals. One way to address these deviations is to augment the linear elasticity model with appropriate surface and interface terms (see e.g.\ Ref.~\cite{Park:07a,Park:08a,Ricci:10a}). An alternative approach is to use molecular dynamics simulations, which automatically account for the effects of surfaces, interfaces and the discrete, atomic nature of matter.

The determination of vibrational properties of complex systems like phononic crystals from molecular dynamics simulations has its own problems. Contrary to crystalline compounds with a small number of atoms per unit cell, the unit cell of two-dimensional hypersonic phononic crystals contains tens of thousands of atoms and simulations of multi-million atom systems are required. This does not only make the simulations computationally expensive. In addition to this, it leads to huge data volumes and it complicates the determination of the vibrational band structure since the signals of many bands must be separated.     

In this work, the vibrational properties of nanoscale phononic crystals are studied with molecular dynamics simulations and finite element method calculations. Results are shown from molecular dynamics simulations of a multi-million atom model phononic crystal system and it is discussed how the technical problems of the determination of the dispersions relations can be overcome. The validity of the resulting vibrational band structure is shown through a comparison with results obtained from finite element method calculations of a simple two-dimensional linear elasticity model. The finite element calculations are further used to answer fundamental questions about the band structure of phononic crystals: In which manner deviate the acoustic modes of the phononic crystal from the plane-waves found in normal bulk materials as the wavelength approaches the Brillouin zone border and how can the lower bands be characterized?

%
% Computational Details
%
\section{Computational Details}
\subsection{Molecular dynamics simulations}\label{SecMD}
\begin{figure}
\centerline{\includegraphics[width=7.5cm]{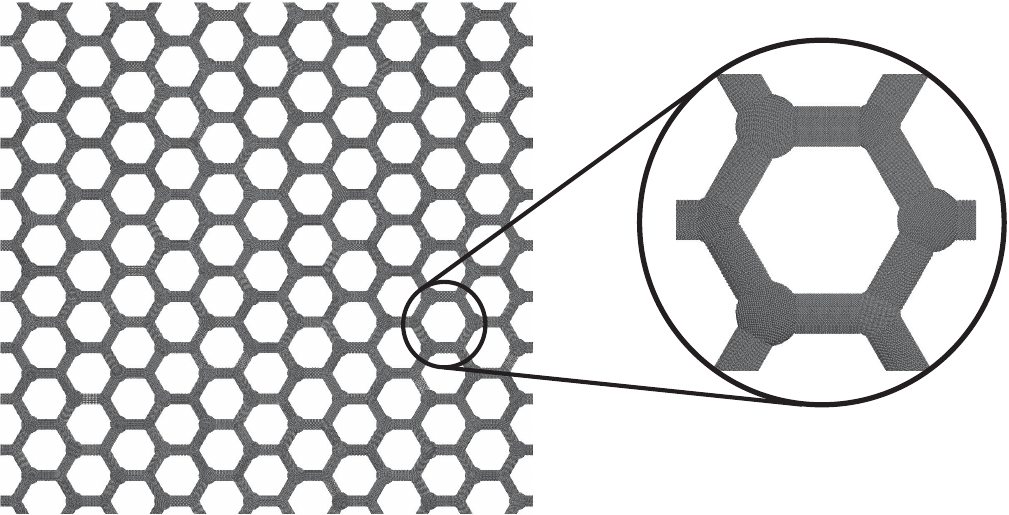}}
\caption{Model phononic crystal used in the simulations}%
\label{FigConf}%
\end{figure}
The molecular dynamics simulations in this work were carried out using the model phononic crystal shown in Fig.~\ref{FigConf}. This model system was created in a bottom-up manner by placing 240 spherical Si nanoparticles with diameters of 7.5\,nm and 9.9\,nm on the sites of a binary honeycomb lattice ($a = 28.87$\,nm) and connecting them with cylindrical nanowires with a diameter of  6\,nm. The axis of the nanowires follows the $[001]$ direction of the Si lattice. In order to enhance the realism of the system, all nanoparticles and nanowires were randomly rotated creating grain boundaries at the interfaces between particles and wires. These boundaries increase the scattering of phonons at the junctions. Furthermore, the different orientations of the constituents implies that the model is not perfectly but only nearly periodic. The model crystal contains 9,009,688 Si atoms and covers an area of $299\,\mathrm{nm} \times 288\,\mathrm{nm}$.

The molecular dynamics simulations of the system employed a modified embedded atom method potential for Si \cite{Baskes:92a}. Although this potential exaggerates the higher phonon frequencies \cite{Heino:07a,Meyer:11a}, it gives a good qualitative description of the phonon spectrum and elastic properties of Si which is sufficient for the purposes of this work. The simulations used the velocity-form of the Verlet algorithm with a time step $\Delta t=1$\,fs. Simulations of systems of this size require the use of parallel computers. Normal domain decomposition, however, has load balancing problems for strongly inhomogeneous systems like the phononic crystal simulated in this work. Therefore the method described in Ref.~\cite{Meyer:13a} was employed to achieve a high parallel efficiency.   

\subsection{Velocity-autocorrelation function}\label{SecVACF}
After a careful equilibration of the model configuration, its vibrational properties were derived with the help of the velocity autocorrelation function. The total vibrational density of states $g(\nu)$ of a system with $N$ atoms is proportional to the Fourier transform of the velocity autocorrelation function  averaged over all atoms:
\begin{equation}
g(\nu) = \int_{-\infty}^{\infty} \mathrm{d}t 
               \frac{\sum_{i=1}^N\left\langle \mathbf{v}_i(t) | \mathbf{v}_i(0)\right\rangle}
                       {\sum_{i=1}^N\left\langle \mathbf{v}_i(0) | \mathbf{v}_i(0)\right\rangle}
               \mathrm{e}^{i\,2\pi \nu t} 
\label{EqVDOS}
\end{equation} 
For periodic systems, frequencies of individual modes can be obtained by projecting the velocities onto a plane wave with wave vector $\mathbf{q}$ and polarization direction $\mathbf{p}$ 
 \begin{equation}
 \hat{v}_\mathbf{q}^\mathbf{p}(t) = \sum_{i=1}^{N} \mathbf{p}\cdot \mathbf{v}_i(t)\,\mathrm{e}^{-i\,\mathbf{q}\cdot\mathbf{r}_i^0}
\label{EqMode}
 \end{equation}
($r_i^0$ denotes the equilibrium position of atom $i$) and calculating the velocity autocorrelation function of the projected velocity \cite{Meyer:11a,Wang:88a}:
\begin{equation}
g_\mathbf{q}^\mathbf{p}(\nu) = \int_{-\infty}^{\infty} \mathrm{d}t 
               \frac{\left\langle {\hat{v}}_\mathbf{q}^\mathbf{p}(t) | \hat{v}_\mathbf{q}^\mathbf{p}(0)\right\rangle}
                       {\left\langle {\hat{v}}_\mathbf{q}^\mathbf{p}(0) | \hat{v}_\mathbf{q}^\mathbf{p}(0)\right\rangle}
               \mathrm{e}^{i\,2\pi\nu t} 
\label{EqVvtQ}
\end{equation} 

The large size of the model phononic crystal creates technical problems for the calculation of the velocity autocorrelation function. Storage of the velocity data of all atoms for a sufficiently large number of time samples would result in more than 12 TB of data. A dataset of this size is not only difficult to store, but  it would also be impractical to process this amount of data in a reasonable time. To overcome this problem a random subset of 250,000 atoms was selected and only the velocity data from these atoms were used in the summations of Eq.~\ref{EqVDOS} and \ref{EqMode}. This sampling method is similar to a Monte-Carlo integration where the integration over the domain is replaced by the summation of a large enough random sample. 

After the selection of the random subset of atoms, the equilibrated model system was simulated over a period of 1.966\,ns. During this simulation, the velocities of the selected atoms were stored at intervals of 15\,fs until $131072 = 2^{17}$ time samples had been collected. The size of this data is "only" 366 GB.   

The velocity data collected during the molecular dynamics simulation were then used to determine the lower bands of the vibrational band structure with the help of the plane wave projected velocity autocorrelation function. For a crystalline system with a simple monoatomic lattice, this task requires the calculation of the function $g_\mathbf{q}^\mathbf{p}(\nu)$ for the desired wave vectors $\mathbf{q}$ with a suitable polarization vector $\mathbf{p}$ to select the band. For example, by choosing $\mathbf{p}$ parallel to $\mathbf{q}$, the resulting function $g_\mathbf{q}^\mathbf{p}(\nu)$ has a peak at the frequency of the longitudinal mode with wave vector $\mathbf{q}$. 

For a complex system with many atoms per unit cell the situation is more complicated since the modes from all bands (more than 225,000 in case of the phononic crystal model system used in this work) contribute to the the function $g_\mathbf{q}^\mathbf{p}(\nu)$. The modes contribute, however, not all in the same way. The strength of the contribution by a mode is determined by the structure of the eigenmode inside the unit cell. If $\mathbf{q} = \mathbf{q}_1 + \mathbf{G}$ where $\mathbf{q}_1$ is a vector in the first Brillouin zone and $\mathbf{G}$ is a reciprocal lattice vector, than the strength of the contribution of the mode is proportional to the absolute square of the $\mathbf{G}$ component of the Fourier series of the function describing the distribution of the vibrational amplitudes.

Two extreme cases are interesting here. In a homogeneous elastic medium, the vibrational modes are simple plane waves. In this case, a given mode will only contribute to the function $g_\mathbf{q}^\mathbf{p}(\nu)$, if the wave vector $q$ matches exactly the wave vector of the mode. Each mode, thus contirbutes only in one specific Brillouin zone. On the other end of the spectrum are crystalline systems with a simple monoatomic lattice. In this case, the amplitude distribution can be described by a single $\delta$-peak in the unit cell and the contributions of the three modes are the same in every Brillouin zone. 

For the determination of the lower bands of the model phononic crystal's band structure, the function $g_\mathbf{q}^\mathbf{p}(\nu)$ was evaluated for multiple equivalent wave vectors in different Brillouin zones. The frequencies of the modes were than determined by fitting Laurentzian functions to those functions $g_\mathbf{q}^\mathbf{p}(\nu)$ which showed the strongest signal for the mode in question. For the polarization vectors $\mathbf{p}$, unit vectors along the x, y, and z directions were used.

%
%  Finite Element Model Description
%
\subsection{Finite element model}
The molecular dynamics simulations were complemented with results from a simple linear elasticity model replicating the honeycomb lattice structure of the molecular dynamics model with two-dimensional slabs of an anisotropic elastic material. While the model replicates the ratio between the nanowire diameter and the lattice constant, no attempt was made to account for the nanoparticles, the varying thickness of the nanowires or the grain boundaries at the junctions between nanoparticles and nanowires.

For simplicity, dimensionless units were employed and the mass density was set to one. The non-zero components of the elasticity tensor were set to:  $c_{xxxx} = c_{yyyy} = 1$, $c_{xxyy} =  0.39$, and $c_{xyxy} = 0.48$. These values replicate the ratios between the elastic constants $C_{11}$, $C_{12}$ and $C_{44}$ of Si \cite{deLaunay:56a}.

The finite element method was used to discretize the anisotropic elastic wave equation for the two-dimensional model systems using a  mesh of 8352 equilateral triangular, linear  $\mathrm{C}^0$  elements per unit cell \cite{Dhatt}. Bloch's theorem was applied, allowing independent computation of the vibrational eigenmodes for each wavevector $\mathbf{q}$.

%
%  Results
%
\begin{figure*}%
\centerline{\includegraphics[height=6cm]{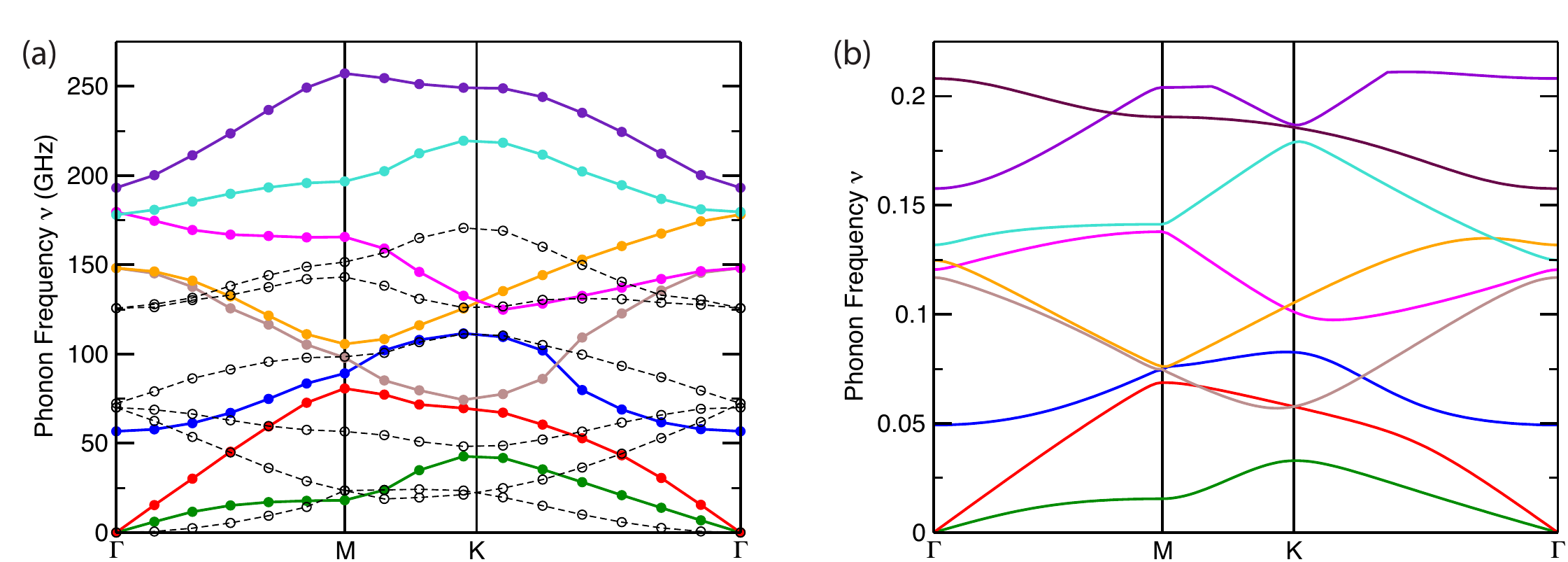}}%
\caption{(Colour online) Phonon band structure of model phononic crystals. (a) Data obtained from molecular dynamics simulations. Circles represent the calculated data points. Lines are only a guide to the eye. Full circles and solid lines show data for in-plane modes whereas open circles and dashed lines belong to out-of-plane (flex) modes. (b) Results of finite element calculations of a two-dimensional linear elasticity model}\label{FigBand}%
\end{figure*}
\section{Results}
\subsection{Vibrational band structure}
The Fig.~\ref{FigBand} (a) shows the band structure obtained from the molecular dynamics simulations of the phononic crystal. The assignment of the data points to bands represented by the same colour is based on the signal strength in different Brillouin zones as well as comparison with Fig.~\ref{FigBand} (b). The modes were classified as in-plane modes with polarization directions in the x-y plane and out-of-plane (flex) modes with a polarization along the z-direction. These flex modes are a result of the fact that we are analyzing the vibrations of an essentially two-dimensional membrane in three dimensions. The nature and dispersion relations of these modes represent an interesting subject on its own which is not pursued in this article. The rest of this work concentrates therefore on the in-plane modes, only.

For the in-plane modes, Fig.~\ref{FigBand} (a) shows the presence of two acoustic bands followed by a complex system of optical bands. In general, it was not possible to assign a longitudinal or transverse character to the bands. The only exception are the two acoustic bands along the $\Gamma$-M direction which are polarized predominantly along the x- and y- directions giving the bands longitudinal acoustic and transverse acoustic character, respectively. In contrast to this, the bands along the $\Gamma$-K direction cannot be characterized as longitudinal or transversal. At the same time, the splitting between the two acoustic bands is much larger at the M point than at the K point of the Brillouin zone.

The band structure calculated from the molecular dynamics simulations shows no sign of a total band gap. However, two partial band gaps can be seen along the $\Gamma$-M  direction. Between 18 and 56 GHz there are no in-plane modes with a transverse polarization along $\Gamma$-M and between 150 and 165 GHz no modes at all exist along this direction. Such partial gaps along certain lattice directions can be used to build acoustic diodes \cite{Maldovan:13a}.

Since the derivation of the band structure from the molecular dynamics simulations is a complicated and delicate process, it is necessary to verify that the results obtained in this manner can be trusted. This verification is given by Fig.~\ref{FigBand} (b) which shows the results obtained from the finite element calculations. Naturally, there are no out-of-plane modes in this figure as the underlying model is purely two-dimensional. 

A comparison between the two panels of Fig.~\ref{FigBand} shows an astonishing agreement between the two calculations, in particular at low frequencies. Not only is there a qualitative agreement between the band structures, but at lower frequencies there is even a quantitative agreement between the ratios of the frequencies at the high symmetry points. This agreement justifies the results from the molecular dynamics simulations.

At higher frequencies there is somewhat less agreement between the two band structures although the general structure remains similar. The molecular dynamics simulations results on the left show less band crossings and the bands seem to repel each other stronger than it is the case for the two-dimensional model.
For example, there is an accumulation of four bands at the $\Gamma$ point near $\nu=0.125$ in the two-dimensional model. The corresponding four bands in the molecular dynamics model are split into to groups around 150 and 180\,GHz (leading to the aforementioned partial gap in this regime). A possible explanation for these differences is the additional mass of the nanoparticles at the lattice points, but more work will be necessary to confirm this hypothesis or find another explanation.

%
%  Mode structure
%
\subsection{Structure of vibrational modes}
\begin{figure*}
\centerline{\includegraphics[width=16.5cm]{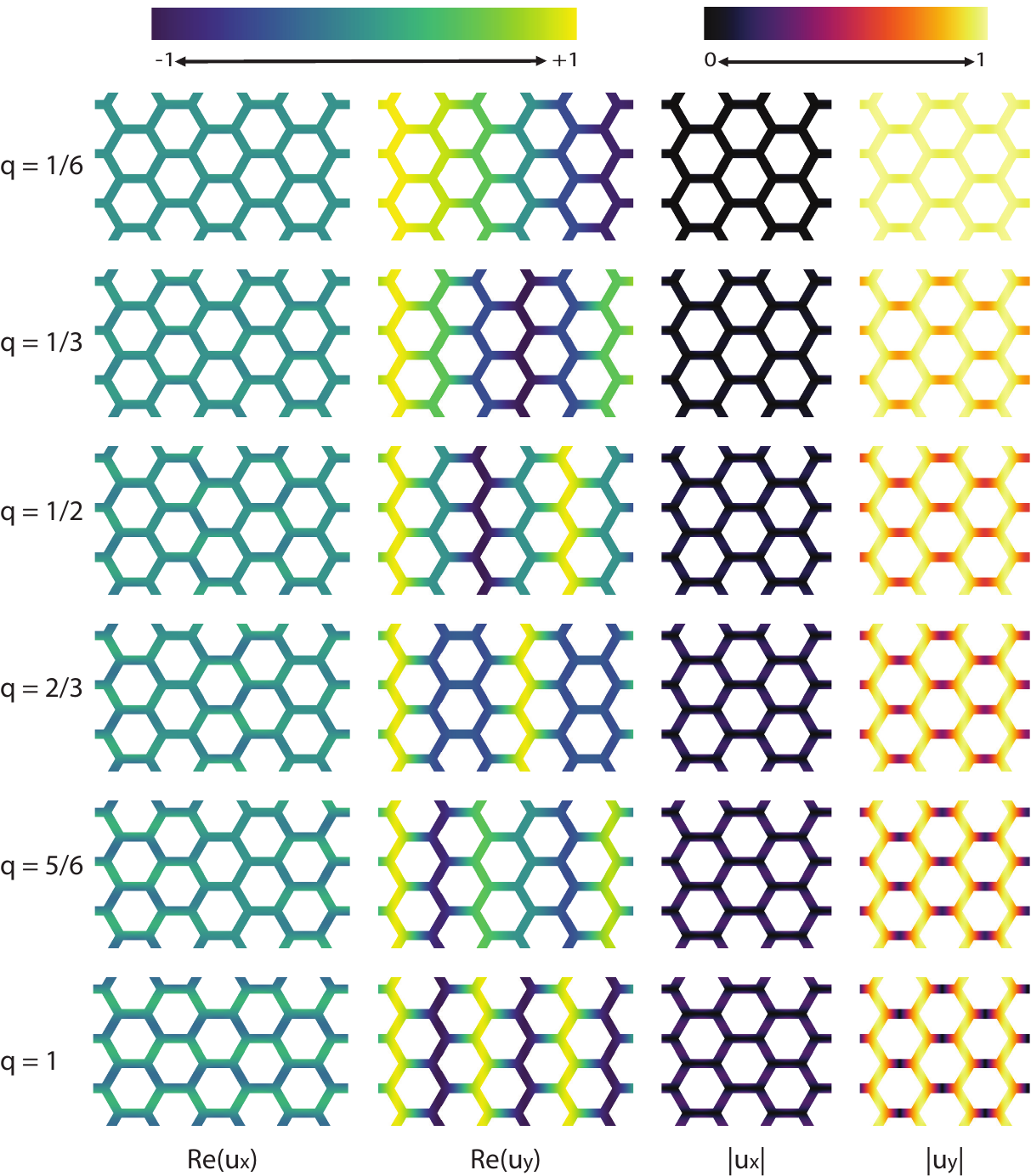}}
\caption{(Colour online) Modes of the transverse acoustic branch along $\Gamma$-M of the finite element model. $\mathbf{q} = \frac{2\pi}{\sqrt{3}a}(q,0)$
}\label{FigTA} 
\end{figure*}
\begin{figure*}
\centerline{\includegraphics[width=16.5cm]{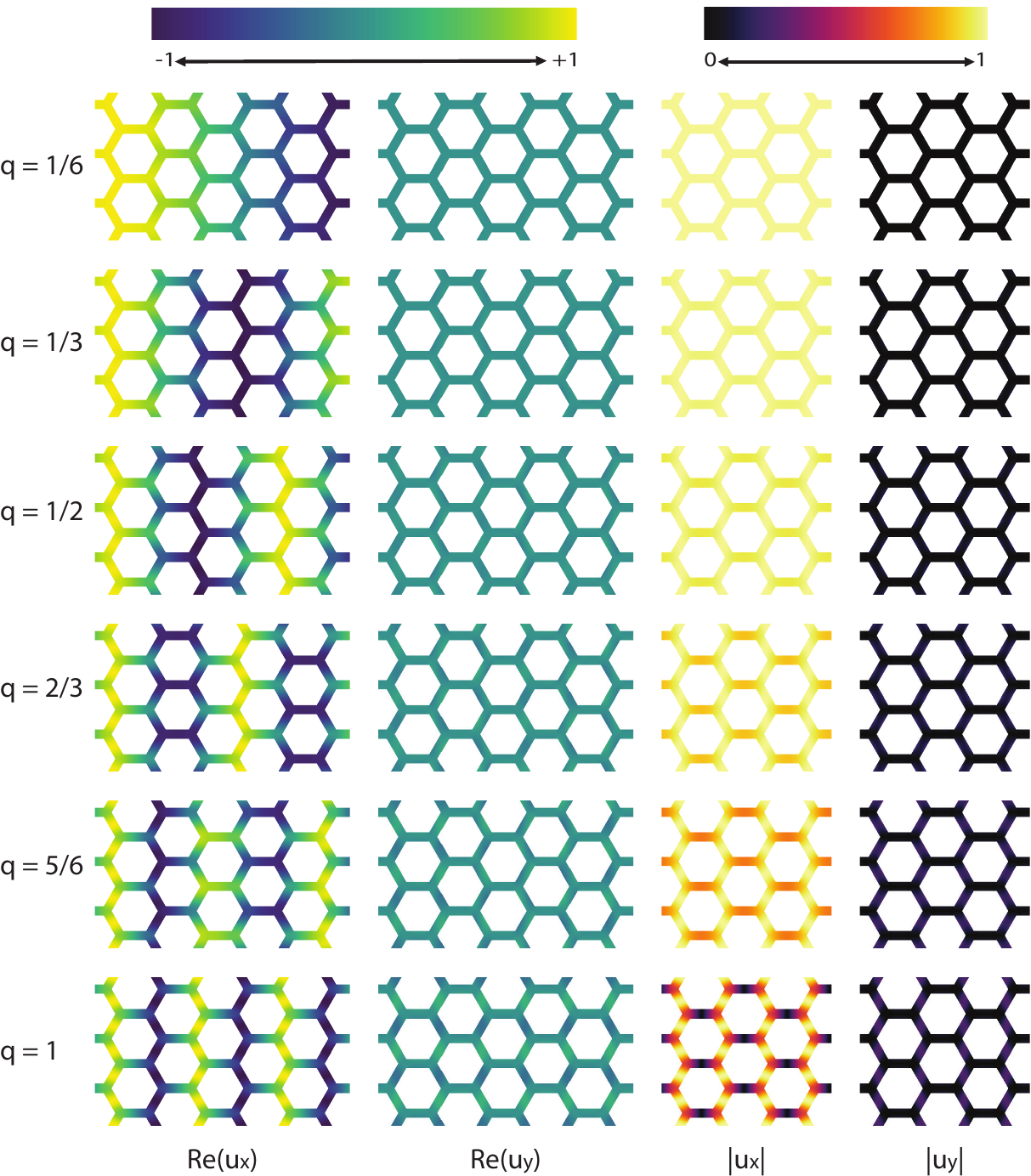}}
\caption{(Colour online) Modes of the longitudinal acoustic branch along $\Gamma$-M of the finite element model. $\mathbf{q} = \frac{2\pi}{\sqrt{3}a}(q,0)$.
}\label{FigLA} 
\end{figure*}
In the preceding section, the two-dimensional finite element model was used to confirm the correctness of the band structure obtained from the molecular dynamics simulations. In this section, the finite element computations are used to study additional questions about the structure of the vibrational modes. Such information is much harder to obtain from molecular dynamics simulations. As discussed in Sec.~\ref{SecVACF}, the relative strength of the velocity autocorrelation function at equivalent wave vectors gives some information about the Fourier series of the amplitude distribution, but this information is incomplete and unsuitable for quantitative comparisons. The finite element computations, on the other hand, not only give the eigenfrequencies but also the complete eigenmodes. 

The vibrational modes of a homogeneous bulk material at wavelengths much larger than the interatomic distances are simple plane waves with a linear dispersion relation. In contrast to this, the band structures discussed in the preceding section show --- as expected--- the occurrence of non-linear dispersion relations in phononic crystals. The band structures do not tell, however, if and in what manner the eigenmodes deviate from the plane wave behaviour of bulk materials. To answer this question, Fig.~\ref{FigTA} and \ref{FigLA} show the structure of several modes in the two acoustic bands along the $\Gamma$-M direction. The left panels of Fig.~\ref{FigTA} and \ref{FigLA} show the real parts of the x- and y-components of the eigenmodes $\mathbf{u}_{\mathbf{q}}(\mathbf{r})$, whereas the panels on the right show the absolute values of the components which correspond to the vibrational amplitudes. The imaginary parts are not shown. Their behaviour is generally similar to the real parts except for a shift by a quarter wavelength. Both figures clearly show the polarization of the bands along the x- and y-directions, making it possible to assign transverse (Fig.~\ref{FigTA}) and longitudinal (Fig.~\ref{FigLA}) characters to the bands (in accordance with the molecular dynamics results).

At first glance, the left panels of Fig.~\ref{FigTA} and \ref{FigLA} suggest that the modes follow the plane wave behaviour. However, the absolute values shown in the right panels make it clear that this is not the case. One of the characteristics of a plane wave is its uniform amplitude which would require constant colours in the right panels of the figures. Instead, the figures show the occurrence of darker spots in the middle of the horizontal nanowires as the wavelength becomes shorter. These darker spots indicate a reduced vibrational amplitude in this region. For $q=1$, i.e. at the Brillouin zone boundary, the centers of the horizontal nanowires practically come to rest and do not participate in the acoustic vibrations.  

\begin{figure*}
\centerline{\includegraphics[width=16.5cm]{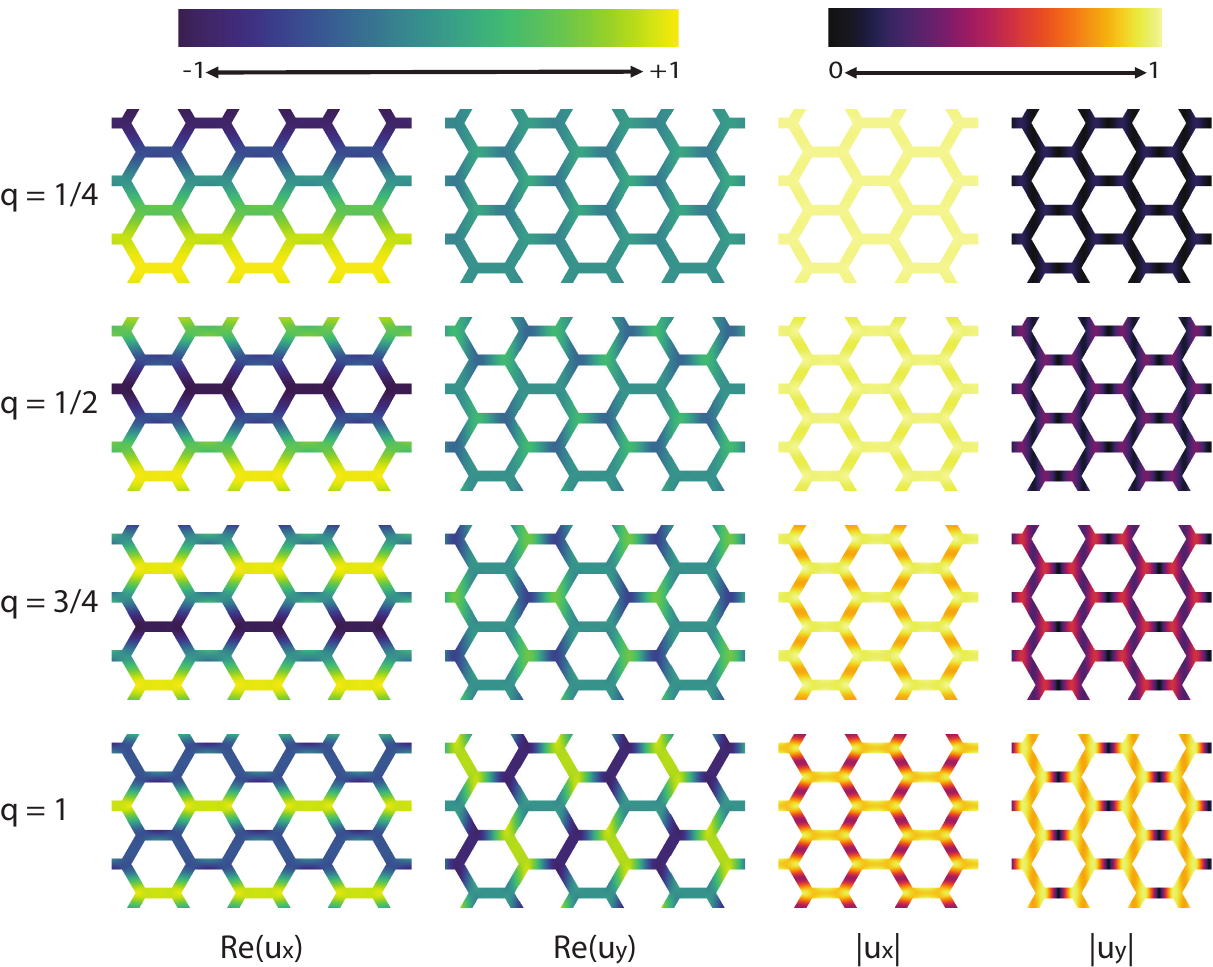}}
\caption{(Colour online) Modes of the lowest acoustic branch along $\Gamma$-K of the finite element model. $\mathbf{q} = \frac{4\pi}{3 a}(0,q)$.
}\label{FigTA2} 
\end{figure*}
A similar behaviour is observed for the modes along the $\Gamma$-K direction shown in Fig.~\ref{FigTA2}. For these upward moving modes, the figure shows the appearance of reduced vibrational amplitudes in the centers of the diagonally oriented nanowires as the wavelength becomes shorter.  In addition to this, a change of the polarization of the modes can be seen. While the first mode ($q = 1/4$) is polarized purely along the x-direction --- i.e. transverse ---, the polarization becomes increasingly mixed for shorter wavelengths. This is in agreement with the molecular dynamics simulations, where it was not possible to assign a polarization character to the acoustic bands along the $\Gamma$-K directions.

Included with this article are three supplemental files with video animations of the evolution of the vibrational modes in the three bands discussed so far. These animations --- which also include the imaginary part of the solution --- demonstrate the changes of the modes as a function of the wavelength in a continuous manner.

Figures~\ref{FigTA}-\ref{FigTA2} show that the acoustic modes deviate from the plane wave behaviour of the bulk through an increasing reduction of the amplitude in the nanowires along the propagation direction of the wave. This reduction imposes a periodic modulation on the plane wave with the periodicity length of the phononic crystal. Indications of this behaviour are also observed in the molecular dynamics simulations. The acoustic modes with the longest wavelengths show strong signals only in the velocity autocorrelation functions calculated in the first Brillouin zone as one would expect for a plane-wave like mode. However, a smaller signal whose strength increases for shorter wavelengths is discernible in the velocity autocorrelation functions calculated in the second Brillouin zone. This indicates that the acoustic modes in the molecular dynamics simulations undergo a similar periodic modulation as shown by Fig.~\ref{FigTA}-\ref{FigTA2} for the finite element model.

\begin{figure*}
\centerline{\includegraphics[width=16.5cm]{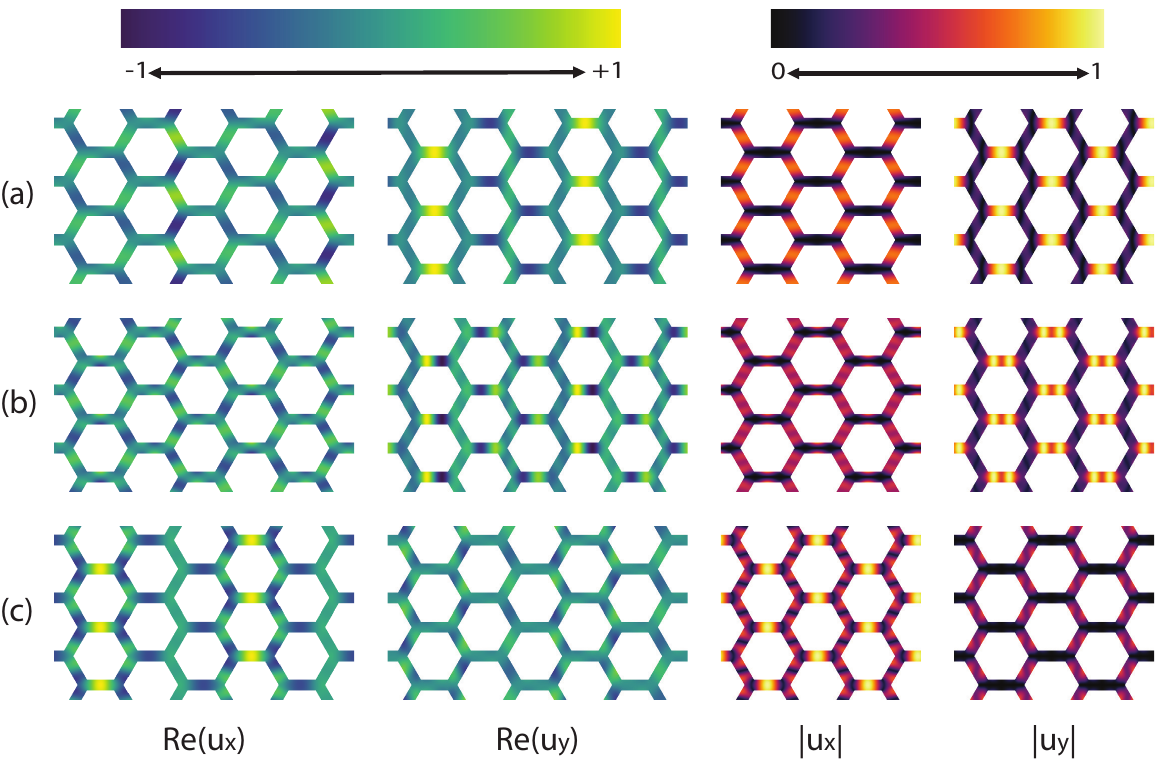}}
\caption{(Colour online) Structure of the third (a), eighth (b) and ninth (e) lowest vibrational mode of the finite model at $\mathbf{q} = \frac{2\pi}{\sqrt{3}a}(\frac{2}{3},0)$.
}\label{FigOptical} 
\end{figure*}
A general discussion of the optical modes of phononic crystals is difficult for two reasons: First, the number of optical bands in phononic crystals is very large. Second, while the acoustic modes must always approach the plane wave behaviour in the long wave-length limit, there is no such restriction for the optical modes. The optical modes are therefore more strongly linked to the geometric details of the system. For these reasons, the discussion of optical modes is restricted to three examples shown in Fig.~\ref{FigOptical}. 

The mode in Fig.~\ref{FigOptical} (a) belongs to the lowest optical band along the $\Gamma$-M direction. The amplitude maximum of this mode lies in the center of the horizontal nanowires with a polarization along the y-direction. This means that this optical mode has its maximum where the vibrations of the corresponding transverse acoustic mode ($q=2/3$ in Fig.~\ref{FigTA}) are reduced. The two modes thus complement each other. This can also be interpreted as a decoupling of the vibrations of the vertical zig-zag chains (which include the nanoparticles in the molecular dynamics model)  and the horizontal nanowires that connect them. At long wavelengths the vibrations of the two are connected. As the wavelength approaches the Brillouin zone boundary, however, the acoustic modes concentrate on the vertical chains while the connecting nanowires are increasingly governed by optical modes with higher frequencies. 

Figures~\ref{FigOptical} (b) and (c) give two more examples of optical modes. The mode in Fig.~\ref{FigOptical} (b) is similar to that in Fig.~\ref{FigOptical} (a) but with two maxima along the nanowires. Both modes have the maximum of their vibrations perpendicular to the nanowires. In contrast to this, Fig.~\ref{FigOptical} (c) is similar to Fig.~\ref{FigOptical} (a) but with the maximum vibration directed along the nanowire.

In general, it was observed that many but not all of the lower optical bands have vibrational maxima along the nanowires and minima at the connection points. These bands can be characterized by the number of vibrational maxima along the nanowires. The movement of the connection points is dominated --- although not exclusively controlled --- by the acoustic bands. One might speculate that this  effect is even more pronounced in the molecular dynamics model where more mass is concentrated in the nanoparticles at the connection points.

%
%  Summary
%
\section{Summary and conclusions}
In this work, a model phononic crystal, built from Si nanoparticles and nanowires, has been studied with large-scale molecular dynamics simulations. Its vibrational band structure was obtained from the simulations with the help of the velocity-projected velocity autocorrelation function. These calculations used a sampling technique to keep the size of the velocity data set manageable.

The band structure obtained from the molecular dynamics simulations is in excellent agreement with results from finite element calculations of a simple two-dimensional model system. This agreement is evidence that it is possible to derive the vibrational band structure of complex multi-million atom systems from molecular dynamics simulations. This is important for nanostructured systems where continuum methods struggle to account for the effects of surfaces, interfaces and the discrete atomic nature of materials. This work shows that molecular dynamics simulations which automatically account for these effects can be used as an alternative or complementary method to study the vibrational properties of complex nanostructures.

In many cases, the computational complexity will prohibit the use of molecular dynamics simulations as a complete replacement of linear-elasticity theory based models. Molecular dynamics simulations can, however, play a vital role to study the impact of nanoscale and atomistic features on vibrational properties. The results can then be used to guide the adaptation of linear-elasticity models to account for nanoscale effects through additional terms and to determine appropriate parameters. Moreover, molecular dynamics simulations give access to a wide range of mechanical  and thermal properties. For example, it is possible to study the vibrational properties and the thermal conductivity of the same microscopic model with molecular dynamics simulations.

The two-dimensional finite element calculations were used further to gain insight into the nature of the vibrational modes of phononic crystals and the character of the lowest bands of the band structure. The results show that as the wavelength approaches the Brillouin zone border, the acoustic modes in the phononic crystal deviate increasingly from the plane-wave behaviour through a periodic modulation with the periodicity length of the phononic crystal. Indications of a similar behaviour are seen in the molecular dynamics simulations. The periodic modulation of the acoustic modes can be interpreted as a manifestation of a decoupling of the vibrations of different parts of the system. Along the [10] direction, this decoupling separates the vibrations of vertical zig-zag chains from the vibrations of the horizontal nanowires connecting them.

The analysis of the band structure of phononic crystal performed in this work is only a beginning and by no means exhaustive. More work is required to obtain a detailed understanding of the nature of band structures of phononic crystal systems. In particular, it would be interesting to see how the band structure depends on the length-to diameter ratio of the nanowires and how the size of nanoparticles at the intersection affects the results. A detailed characterization of the lower optical bands in phononic crystal band structures would be useful. 

Note: After the completion of this manuscript, work has begun on calculations using an improved finite-element method model. Preliminary results from this yet incomplete work suggest a slightly different assignment of bands to the molecular dynamics results which would lead to a  very small absolute band gap around 100\,GHz.

%
%  Acknowledgments
%
\begin{acknowledgement}%
The author thanks Abdellatif Serghini Mounim for helpful discussions about the application of Floquet theory infinite element calculations. This work has been supported financially by Laurentian University and  the Natural Sciences and Engineering Research Council of Canada (NSERC). Generous allocation of computer time on the facilities of Compute/Calcul Canada is gratefully acknowledged. Figures~\ref{FigTA}-\ref{FigOptical} use the colourmaps \textit{viridis} and \textit{inferno} from Ref.~\cite{matplotlib}.
\end{acknowledgement}

%
% References
%
%\bibliography{phoncrys}
\providecommand{\WileyBibTextsc}{}
\let\textsc\WileyBibTextsc
\providecommand{\othercit}{}
\providecommand{\jr}[1]{#1}
\providecommand{\etal}{~et~al.}

\end{document}